\documentclass[fleqn,11pt,twoside]{article}

\usepackage{amsthm,amsthm,amssymb, color, xcolor,epsfig, graphics, subfigure}

\usepackage{amsmath, graphicx, latexsym, lscape }

\usepackage{amsmath,amssymb}
\usepackage{mathrsfs}
\usepackage{amsthm}
\usepackage{amsfonts}
\usepackage{amsmath,amsfonts,amssymb,amsthm,url,array}
\usepackage{bbm}
\usepackage{bm}

\makeatletter
\newcommand{\copyrightnote}[2]{{\renewcommand{\thefootnote}{}
 \footnotetext{\small\it
\begin{flushleft}
 \copyright \ #1   #2  
\end{flushleft}}}}

\newcommand{\Name}[1]{\begin{flushleft}
                       \LARGE \bf #1
                       \end{flushleft}\vspace{-3mm}}

\newcommand{\Author}[1]{\begin{flushleft}
                       \it #1 \end{flushleft}}

\newcommand{\Address}[1]{\begin{flushleft}
                       \it #1 \end{flushleft}}

\newcommand{\Date}[1]{\begin{flushleft}
                      \small  \it #1 \end{flushleft}}

\newcommand{\evenhead}{Author \ name}
\newcommand{\oddhead}{Article \ name}

\renewcommand{\@evenhead}{
\hspace*{-3pt}\raisebox{-15pt}[\headheight][0pt]{\vbox{\hbox to \textwidth
{\thepage \hfil \evenhead}\vskip4pt \hrule}}}
\renewcommand{\@oddhead}{
\hspace*{-3pt}\raisebox{-15pt}[\headheight][0pt]{\vbox{\hbox to \textwidth
{\oddhead \hfil \thepage}\vskip4pt\hrule}}}
\renewcommand{\@evenfoot}{}
\renewcommand{\@oddfoot}{}

\newcommand{\q}[1]{``#1''}
\newlength{\tagwidth}
\newcommand{\TAG}[1]{\tag*{$\parbox{\tagwidth}{#1}$}}

\long\def\@makecaption#1#2{%
  \vskip\abovecaptionskip
  \sbox\@tempboxa{\small \textbf{#1.}\ \ #2}%
  \ifdim \wd\@tempboxa >\hsize
    {\small \textbf{#1.}\ \ #2}\par
  \else
    \global \@minipagefalse
    \hb@xt@\hsize{\hfil\box\@tempboxa\hfil}%
  \fi
  \vskip\belowcaptionskip}

\newcommand{\JNMPnumberwithin}[3][\arabic]{%
  \@ifundefined{c@#2}{\@nocounterr{#2}}{%
    \@ifundefined{c@#3}{\@nocnterr{#3}}{%
      \@addtoreset{#2}{#3}%
      \@xp\xdef\csname the#2\endcsname{%
        \@xp\@nx\csname the#3\endcsname .\@nx#1{#2}}}}%
}

\newcommand{\resetfootnoterule} {
  \renewcommand\footnoterule{%
  \kern-3\p@
  \hrule\@width.4\columnwidth
  \kern2.6\p@}
}

\renewcommand{\footnoterule}{}

\makeatother

\theoremstyle{definition}

\begin{document}

\renewcommand{\evenhead}{ {\LARGE\textcolor{blue!10!black!40!green}{{\sf \ \ \ ]ocnmp[}}}\strut\hfill A.S. Fokas and A. Latifi}
\renewcommand{\oddhead}{ {\LARGE\textcolor{blue!10!black!40!green}{{\sf ]ocnmp[}}}\ \ \ \ \   The Nonlinear Schr\"odinger equation with forcing}

\thispagestyle{empty}
\newcommand{\FistPageHead}[3]{
\begin{flushleft}
\raisebox{8mm}[0pt][0pt]
{\footnotesize \sf
\parbox{150mm}{{Open Communications in Nonlinear Mathematical Physics}\ \  \ {\LARGE\textcolor{blue!10!black!40!green}{]ocnmp[}}
\ \ Vol.2 (2022) pp
#2\hfill {\sc #3}}}\vspace{-13mm}
\end{flushleft}}

\FistPageHead{1}{\pageref{firstpage}--\pageref{lastpage}}{ \ \ Article}

\strut\hfill

\strut\hfill

\copyrightnote{A.S. Fokas and A. Latifi. Distributed under a Creative Commons Attribution 4.0 International License}

\Name{The Nonlinear Schr\"odinger equation with forcing involving products of eigenfunctions}

\Author{A.S. Fokas$^{\,1,\,2}$ and A. Latifi$^{\,3}$}

\Address{$^{1}$ 	Department of Applied Mathematics and Theoretical Physics, 
	University of Cambridge, 
	Cambridge CB3 0WA, UK \\[2mm]
$^{2}$ Viterbi School of Engineering, 
University of Southern California, Los Angeles, 
California, 90089, USA \\[2mm]
$^{3}$ 	Department of Mechanics, Faculty of Physics, 
Qom University of Technology, 
Qom, Iran }

\Date{Received July 19 2022; Accepted July 31 2022}

\setcounter{equation}{0}

\begin{abstract}
\noindent 
We elaborate on a new methodology, which starting with an integrable evolution equation in one spatial dimension, constructs an integrable {\it forced} version of this equation. The forcing consists of terms involving quadratic products of certain eigenfunctions of the associated Lax pair. Remarkably, some of these forced equations arise in the modelling of important physical phenomena. The initial value problem of these equations can be formulated as a Riemann-Hilbert problem, where the \q{jump matrix} has explicit $x$ and $t$ dependence and can be computed in terms of the initial data. Thus, these equations can be solved as efficiently as the nonlinear integrable equations from which they are generated. Details are given for the forced versions of the nonlinear Schr\"odinger.
\end{abstract}

\label{firstpage}


\section{Introduction}

Applied complex analysis has had a crucial impact on the theory of a particular class of nonlinear equations, called integrable.  In particular, the formulation and solution of a fundamental problem arising in complex variables, called a Riemann-Hilbert (RH) problem, is at the heart of various approaches for solving integrable nonlinear evolution PDEs in one spatial dimension. One such powerful approach is
the so-called \q{dressing method}, which both constructs and solves   integrable nonlinear evolution equations starting with a given RH problem. This technique, which has its origin in the pioneering works of Zakharov and Shabat \cite{ZS}, starts with postulating a  RH problem  \cite{Ablowitz_Fokas}, formulated as follows: let the $2\times 2$ matrix-valued function $\mu(x,t,k)$ be analytic for all values of the complex variable $k$ except for $k$ real, where it has a \q{jump}. Specifically,
\begin{subequations}\label{RH problem}
	\begin{equation}\label{RH problem1}
		\mu^-(x,t,k)=\mu^+(x,k,t)J(x,t,k),\qquad k\in \mathbb{ R},
	\end{equation}
	where $\mu^+(x,t,k)$ and $\mu^-(x,t,k)$ denote the limits of $\mu$ as $k$ approaches the real axis from above and below, respectively. Given the  $2\times 2$ matrix $J(x,t,k)$, determine $\mu$, under the assumption 
	\begin{equation}\label{RH problem2}
		\mu(x,t,k)=I + O\left(\frac{1}{k}\right), \qquad \left | k\right | \rightarrow \infty,
	\end{equation}
\end{subequations}
where $I$ denotes the  $2\times 2 $ identity matrix.\\

The algorithmic determination of a nonlinear equation that can be solved via \eqref{RH problem} involves two steps: (i) The identification of two operators, $ M_x$ and $ M_t$, by the requirement that they \q{commute} with the jump matrix $J$, which means that the functions $ M_x\mu$ and $ M_t\mu$, satisfy the same jump condition as $\mu$. This requirement fixes the $x$ and $t$ dependence of the matrix $J$. (ii) The imposition of appropriate relationships between the functions defining the operators  $ M_x$ and $ M_t$ by the requirement that $ M_x\mu$ and $ M_t\mu$, are of order $1/k$  for large  $k$. The conditions (i) and (ii), imply that $ M_x\mu$ and $ M_t\mu$, satisfy the homogeneous version of the RH problem \eqref{RH problem}. Then,  the {\it assumption} that the latter problem has only the zero solution, implies $ M_x\mu= M_t\mu=0$. These two equations define a Lax pair associated with the RH problem \eqref{RH problem}. The terms of the $t$-part of the Lax pair which are of order $1/k$ for large $k$ yield a nonlinear PDE that can be solved via \eqref{RH problem}.\\

It is straightforward to postulate a RH problem for an $ N \times N $ matrix with a jump along a contour $ L $ instead of the real axis. However, for the results of this paper it is sufficient to consider the RH problem defined by Eqs \eqref{RH problem}.\\

\noindent {\underline{The new methodology}}\\

Starting with a given integrable PDE, the new methodology constructs an integrable  forced version of this PDE by modifying the operator $ M_t$. This modification involves the operator  $ \Delta_g $ defined as follows:
\begin{align}\label{Deltag}
	(\Delta_gf)&(x,t,k)=-\left(\dfrac{1}{2\mathrm{i}\pi}\int_{ \mathbb{ R}}\frac{g(t,l)}{l-k}dl\right)f(x,t,k)\sigma_3 \notag\\
	&+\left(\dfrac{1}{2\mathrm{i}\pi}\int_{ \mathbb{ R}}\frac{g(t,l)}{l-k}f(x,t,l)\sigma_3\left(f(x,t,l)\right)^{-1}dl\right)f(x,t,k),\quad k\in \mathbb{C},
\end{align}
where $ \sigma_3 $ is the third Pauli matrix, namely, 
\begin{equation}\label{Pauli}
	\sigma_3=\text{diag}(1,-1),
\end{equation}
and $ g(t,k) $ is a scalar real function, which also satisfies an additional constraint that will be discussed in section 2. The definition of $ \Delta_gf $ implies that this operator is well defined for any function $ f $ which is sectionally analytic with respect to the real axis. Namely, let $ f^{+} $ and $ f^{-} $ denote the values of $ f $ for $ \text{Im}k>0 $ and $ \text{Im}k<0 $, respectively. Then, \eqref{Deltag} becomes
\begin{align}\label{Deltag clarification}
	(\Delta_g&f^{\pm})(x,t,k)=-\left(\dfrac{1}{2\mathrm{i}\pi}\int_{ \mathbb{ R}}\frac{g(t,l)}{l-k}dl\right)f^{\pm}(x,t,k)\sigma_3\notag \\
	&	+\left(\dfrac{1}{2\mathrm{i}\pi}\int_{ \mathbb{ R}}\frac{g(t,l)}{l-k}f^{\pm}(x,t,l)\sigma_3\left(f^{\pm}(x,t,l)\right)^{-1}dl\right)f^{\pm}(x,t,k),\quad \text{Im}k\gtrless 0. 
	\TAG{$ (4)^{\pm} $}
\end{align}
\refstepcounter{equation}
Starting with the above definitions, it is straightforward to compute the limits of equations \ref{Deltag clarification} as $\text{Im} k\to 0 $. For this purpose we use the so-called Plemelj's formulas \cite{Ablowitz_Fokas},
\begin{equation*}\label{Plemelj}
	\lim_{\epsilon \rightarrow 0}\frac{1}{2\mathrm{i}\pi}	\int_{ \mathbb{ R}}\frac{g(t,l)dl}{l-(k\pm \mathrm{i}\epsilon)}=\pm \frac{g(t,k)}{2}+\frac{1}{2\mathrm{i}\pi}p\int_\mathbb{R}\frac{g(t,l)dl}{l -k}, \qquad k\in \mathbb{ R},
\end{equation*}
where the symbol $ p $ denotes the principal valued integral.\\

Using these formulas, Equations \ref{Deltag clarification} yield,
\begin{align}\label{pm}
	(\Delta_g&f^{\pm})(x,t,k)=-\left(\dfrac{1}{2\mathrm{i}\pi}p\int_{ \mathbb{ R}}\frac{g(t,l)}{l-k}dl\right)f^{\pm}(x,t,k)\sigma_3+ \nonumber\\
	&\left(\dfrac{1}{2\mathrm{i}\pi}p\int_{ \mathbb{ R}}\frac{g(t,l)}{l-k}f^{\pm}(x,t,l)\sigma_3\left(f^{\pm}(x,t,l)\right)^{-1}dl\right)f^{\pm}(x,t,k),\quad k\in \mathbb{R}.
\end{align}
Equation \eqref{pm} shows that the operator $ \Delta_g $ has the crucial property that  $ \Delta_gf^{+} $ and $ \Delta_gf^{-}$ satisfy the same equation.
Indeed, the contribution of the terms which do not involve the principal valued integrals cancel:
\begin{equation*}\label{canceled}
	\pm \frac{g}{2}f^{ \pm }\sigma_3 \pm \frac{g}{2}f^{\pm}\sigma_3(f^{\pm})^{-1}f^{\pm}=0.
\end{equation*}

It is shown in section 2 that despite the above simplification, the operator $ \Delta_g $ does \textit{not} commute with $ J $. Hence, it is\textit{ not} possible to use the dressing method in order to obtain the time dependence of the  matrix $ J $, \textit{nor} to conclude that the equation $ M_t\mu=0 $ involving $ \Delta_g\mu $ is compatible with the equation $ M_x\mu=0 $.\\

The novelty of the new methodology is that it goes beyond the dressing method. This methodology involves the algorithmic steps described below.

\begin{description}
	\item[(i)] Modify the $ t $-part of the Lax pair of a given integrable PDE by including the term $ \Delta_g $. In some cases, the inclusion of this term necessitates the introduction of additional, local terms. This is further discussed in \cite{FOKAS2022128290}.
	\item[(ii)] Using the $ x $-part of the Lax pair (which is unchanged) and implementing the usual methodology of the inverse scattering method, formulate the RH problem \eqref{RH problem} for the eigenfunctions $ \mu^{+} $ and $ \mu^{-} $. The jump matrix $ J $ has an explicit $ x $-dependence but its $ t $-dependence needs to be determined.
	\item[(iii)] Using the new $ t $-part of the Lax pair, compute the $ t $-dependence of the jump matrix $ J $. Remarkably, despite of the occurrence of certain nonlinear terms in $ M_t $, the time evolution of $ J $  can be written \textit{explicitly}. 
	\item[(iv)] Obtain an integrable forced PDE by requiring that the terms of $ M_t\mu=0 $, which are of order $ 1/k $, $ k $ large, vanish. In this connection it is noted that the expansion of $ \mu $ is determined by analysing the large $ k $ asymptotics of the expansion  $ M_x\mu=0 $ and by requiring that the terms of  $ M_t\mu=0 $ which are of order $ k $ and order $ 1 $, $ k $ large, also vanish.
	\item[(v)]  Prove that the solution of  RH problem \eqref{RH problem} satisfies both parts of the Lax pair; this implies that this solution can be used to solve the forced PDE.  For the unforced PDE, this step relies on using the dressing method, which as stated earlier, is {\it not} applicable to the new $ M_t $. However, it still possible to prove that the solution of the RH problem does satisfy the Lax pair, by using the so-called direct linearizing transform (DLT), which was introduced by one of the authors and Mark Ablowitz exactly 40 years ago.  Indeed, the RH problem \eqref{RH problem} gives rise to a linear integral equation; starting with this equation, it can be shown explicitly that the solution of this integral equation satisfies the new $ M_t $. It should be emphasized that the DLT  provides a unique way of choosing the correct matrix eigenfunctions appearing in the forcing of the new integrable PDE: it is the eigenfunction appearing in the integral equation of the DLT, which arises from the given RH problem; hence,  it is $\mu^+$.
\end{description}


Incidentally,  the DLT starts with a more general integral equation than the one arising from the given RH problem; it starts with an equation involving an arbitrary measure and contour. Thus, the DLT can capture more general solutions than those obtained via the inverse scattering. Actually, as first understood by Ralph Coifman \cite{coifman}, the DLT provides the generalization from linear to integrable PDEs of the celebrated \q{Ehrenpreis fundamental principle}. For linear evolution PDEs with constant coefficients, this result states that for a smooth, bounded, and convex domain, the solution of a well posed problem can be expressed in the form that involves the same $x$ and $t$ dependence as the Fourier transform representation of the solution of the initial value problem. For example, for the linearized NLS, this result implies that there exists an appropriate measure, $d\rho(l)$, and contour, ${\mathscr L}$, such that the solution of a well posed problem can be expressed in the form

\begin{equation*}\label{measure rho}
	u(x,t)=\int_{\mathscr L}\mathrm{e}^{\mathrm{i}lx+\mathrm{i}l^2t}d\rho(l).
\end{equation*}\\

\noindent \underline{Notation}\\

$ C{f}(k) $ will denote the Cauchy integral
\begin{equation}\label{Cauchy}
	C{f}(k)=\frac{1}{2\mathrm{i}\pi}\int_{\mathbb{R}}\frac{f(l)}{l-k}dl, \qquad k\in \mathbb{C}.
\end{equation}	

Then, eq.\eqref{Deltag} becomes
\begin{equation}
	\Delta_{g}f=-(C{g})f\sigma_3+(Cgf\sigma_3f^{-1})f, \qquad k\in \mathbb{C}.
\end{equation}
$ Hf $ will denote the Hilbert transform
\begin{equation}
	(Hf)(k)=\frac{1}{\pi}p\int_{\mathbb{ R}}\frac{f(l)}{l-k}dl, \qquad k\in \mathbb{R}.
\end{equation}
Then, Eqs \eqref{pm} become
\begin{equation}
	\Delta_{g}f^{\pm}=-\frac{1}{2\mathrm{i}}(Hg)f^{\pm}\sigma_3+\frac{1}{2\mathrm{i}}\left(Hgf^{\pm}\sigma_3(f^{\pm})^{-1}\right)f^{\pm}, \qquad k\in \mathbb{R}.
\end{equation}\\

\noindent \underline{Organization of the Paper and Main Results}\\

In Section 2, the new methodology is implemented to the NLS. It shows that the following forced version of NLS is integrable:
\begin{align}\label{full NLS}
	\mathrm{i}q_t+\frac{\alpha}{2}q_{xx}-\alpha\lambda|q|^2q=\frac{2\mathrm{i}}{\pi}\int_{\mathbb{R}}\frac{g(t,k)}{a_0(k)}\Phi_1^+(x,t,k)\Psi^+_1(x,t,k)&\mathrm{e}^{\mathrm{i}\lambda \left(HG(t,k)|b_0|^2(k)\right)}dk,\nonumber\\
	& x\in\mathbb{R},\quad t>0,
\end{align}
where $\alpha$ is a constant real parameter, $\lambda=\pm1$, $\Phi^+_1$, $\Psi^+_1$ are appropriate eigenfunctions specified in \eqref{phi-psi}-\eqref{limit},
\begin{align}\label{mathematical boundary conditions}
	G(t,k)=\int_0^tg(\tau,k)d\tau,\quad &b_0(k)=\lim_{x\rightarrow -\infty}\mathrm{e}^{2\mathrm{i}kx}\Psi^+_{10}(x,k),
	\nonumber \\
	&a_0(k)=\lim_{x\rightarrow -\infty}\Psi^+_{10}(x,k),
\end{align}
and $(\Psi_{10}^+(x,k), \Psi_{20}^+(x,k))$ are defined in terms
of $q_0(x)=q(x,0)$ by
\begin{align}\label{voltera}
	&\Psi_{10}^+(x,k)= -\int_x^\infty  d\xi \mathrm{e}^{2\mathrm{i}k(\xi-x)}q_0(\xi) \Psi_{20}^+(\xi,k), \nonumber\\ 
	&\Psi_{20}^+(x,k)=1-\int_x^\infty d\xi \lambda\overline{ q}_0(\xi) \Psi_{10}^+(\xi,k),
	\qquad -\infty<x<\infty, \quad \text {Im}k\ge 0.
\end{align}
Eq. \eqref{full NLS} possesses the following Lax pair:
\begin{equation}\label{x evolution of X}
	X_x+\mathrm{i}k[\sigma_3,X]-QX=0,\qquad
	Q=\begin{pmatrix} 0 &  q \cr \lambda\overline{q} & 0 \end{pmatrix},
\end{equation}
\begin{align}\label{t evolution of X}
	X_t+&\mathrm{i}\alpha k^2[\sigma_3,X]-\alpha\left(kQ-\frac{\mathrm{i}}{2}(Q_x+Q^2)\sigma_3\right)X
	\nonumber\\
	&=\frac{1}{2\mathrm{i}}(Hg)X\sigma_3-\frac{1}{2\mathrm{i}}(HgX\sigma_3X^{-1})X,
	\qquad x\in\mathbbm{R}, \quad t>0, \quad k\in\mathbbm{R}.
\end{align}\\

Assuming that $ a_0(k) $ has no zeros in the closed upper half $ k $-plane, the initial value problem of \eqref{full NLS} can be solved via the RH problem \eqref{RH problem} where the jump matrix $ J $ is given by
\begin{equation}\label{mu matriciel}
	J(x,t,k)=
	\begin{pmatrix} 1 &  -\frac{{b_0(k)}}{\overline{a_0}(k)}\mathrm{e}^{-\mathrm{i}\left[2kx-2k^2t-(H\int_0^tg(\tau,k)d\tau) \right]} \cr\lambda\frac{\overline{b_0(k)}}{a_0(k)}\mathrm{e}^{\mathrm{i}\left[2kx-2k^2t-(H\int_0^tg(\tau,k)d\tau) \right]}  & \frac{1}{|a_0(k)|^2} \end{pmatrix}.
\end{equation}

The DLT, motivated from the linear integral equation solving the above RH  problem, is implemented in Section 3. To minimize computations, we give the details of the proof only for the case that $ \alpha=0 $ and $ g $ is independent of $ t $. It is straightforward to extend the proof when $ g_t\neq 0 $ and $\alpha\neq 0 $.\\

Specific physical applications are presented in Section 4. Our results are further discussed in Section 5. The direct verification that the compatibility condition of the new Lax pairs yield the forced version of NLS is presented in the Appendix A. Solitons are discussed in the Appendix B.

\section{An Integrable Forced Version of the NLS }\label{Lax pair derivation}
In this section we will construct an integrable forced version of the celebrated NLS equation
\begin{equation}\label{NLS}
	\mathrm{i} q_{_t}+\frac{1}{2}q_{_{xx}}-\lambda \left|q\right|^2q=0, \qquad \lambda=\pm 1.
\end{equation}
In order to motivate the construction of the Lax pair of the NLS, we pose the following question:  can we construct an integrable version of the linear equation below?
\begin{equation}\label{LS}
	\mathrm{i}u_t+u_{xx}=0.
\end{equation}
It is important to note that, as emphasized by Israel Gelfand and one of the authors \cite{Fokas1994IntegrabilityOL}, linear equations do possess a Lax pair formulation (this fact was missed by the classical mathematicians of the 18\textsuperscript{th} century, as well as all other investigators of PDEs). A Lax pair of \eqref{LS} is given by
\begin{align}\label{linear lax pair}
	&\varphi_x+\mathrm{i}k\varphi=u, \nonumber \\
	&\varphi_t+\mathrm{i}k^2\varphi=ku+\mathrm{i}u_x, \qquad k\in \mathbbm{C},
\end{align}
where $\varphi$ is a scalar function of $x$, $t$, $k$. The derivation of \eqref{linear lax pair} illustrates the importance of the notion of the \lq\lq{}adjoint\rq\rq{}. Indeed, the formal adjoint of \eqref{LS} is the equation
\begin{equation}\label{adjoint}
	-\mathrm{i}v_t+v_{xx}=0.
\end{equation}
Eqs \eqref{LS} and \eqref{adjoint} imply
\begin{equation}\label{uv}
	\mathrm{i}\left(uv\right)_t=\left(uv_x-vu_x\right)_x.
\end{equation}
Hence, there exists a function $\varphi$, such that 
\begin{align}\label{vphi}
	&(\mathrm{i}v\varphi)_x=uv, \nonumber \\
	&(\mathrm{i}v\varphi)_t=uv_x-vu_x.
\end{align}
Choosing for $v$ the particular solution of \eqref{adjoint} given by $\exp(\mathrm{i}kx+\mathrm{i}k^2t$), $k$ arbitrary complex constant, we find the Lax pair \eqref{linear lax pair}.\\

Since, the nonlinear version of \eqref{LS} may also involve the complex conjugate of $u$, we consider the Lax pair of the system consisting of Eq \eqref{LS} and the equation satisfied by $\displaystyle{{\bar u}}$. This gives rise to the operators 
\begin{equation}\label{x and t operators}
	\partial_x+\mathrm{i}k\sigma_3, \qquad \partial_t+\mathrm{i}k^2\sigma_3, \qquad k\in\mathbbm{C},
\end{equation}
with $\sigma_3$ defined in \eqref{Pauli}.

For the implementation of the first step of the dressing method it is more convenient to introduce the operators $D_x$ and $D_t$ defined by
\begin{equation}\label{D operators}
	D_xf=f_x-\mathrm{i}kf\sigma_3, \qquad D_tf=f_t-\mathrm{i}k^2f\sigma_3.
\end{equation}
These operators commute with the matrix $J$ appearing in \eqref{RH problem} provided that the $x$ and $t$ dependence of $J$ is determined by the equations 
\begin{equation}\label{commutations}
	J_x+\mathrm{i}k[\sigma_3,J]=0, \qquad J_t+\mathrm{i}k^2[\sigma_3,J]=0.
\end{equation}
Indeed
\begin{align}
	D_x\mu^-&= \mu^-_x-\mathrm{i}k\mu^-\sigma_3= (\mu^+J)_x-\mathrm{i}k\mu^+J\sigma_3 \nonumber\\
	&= \mu^+_xJ+\mu^+J_x-\mathrm{i}k\mu^+J\sigma_3=(D_x\mu^+)J,
\end{align}
where for deriving the last equality, we replaced $J_x$ via the first of the equations \eqref{commutations}. Similarly, for $D_t$.\\

Taking into consideration the equality
\begin{equation}\label{A2}
	(f\mu^-)=(f\mu^+)J, \nonumber
\end{equation}
where $f$ is any function which is analytic in $k$, it follows that the operator $M_x \mu$ defined by
\begin{equation}\label{M1mu}
	M_x \mu=\mu_x+\mathrm{i}k[\sigma_3,\mu]-Q(x,t)\mu,
\end{equation}
satisfies \eqref{RH problem1}. Substituting the asymptotic expansion
\begin{equation}\label{expansion}
	\mu=I+\frac{\mu_1(x,t)}{k}+\frac{\mu_2(x,t)}{k^2}+\frac{\mu_3(x,t)}{k^3}+O(\frac{1}{k^4}),\qquad k\rightarrow \infty,
\end{equation}
in Eq \eqref{M1mu}, it follows that $M_x\mu$ is of order $1/k$, for large $k$, provided that $Q$ is the off-diagonal matrix given by
\begin{equation}\label{Q}
	Q=\mathrm{i}[\sigma_3,\mu_1].
\end{equation}
Similarly, the function $M_t \mu$ defined by
\begin{equation}\label{M2mu}
	M_t \mu=\mu_t+\mathrm{i}k^2[\sigma_3,\mu]-kA(x,t)\mu-B(x,t)\mu,
\end{equation}
satisfies \eqref{RH problem1}. Substituting the asymptotic expansion \eqref{expansion} in \eqref{M2mu}, it follows that the terms of order $k$ (for large $k$) of $M_t \mu$ vanish, provided that
\begin{equation}\label{defiinition of A}
	A=\mathrm{i}[\sigma_3,\mu_1]=Q,
\end{equation}
where in the second equality above, we used the identity \eqref{Q}. The terms  of order 1 of $M_t \mu$ vanish provided that
\begin{equation}\label{defiinition of B}
	B+Q\mu_1=\mathrm{i}[\sigma_3,\mu_2].
\end{equation}
Thus,
\begin{equation}\label{BD-BO}
	B_D=-Q\mu_{1O},\qquad
	B_O+Q\mu_{1D}=\mathrm{i}[\sigma_3,\mu_{2O}],
\end{equation}
where the subscripts $D$ and $O$ refer to the diagonal and off diagonal parts of the given matrix, respectively.\\
Equation \eqref{Q} implies
\begin{equation}\label{mu1O}
	\mu_{1O}=\frac{\mathrm{i}}{2}Q\sigma_3.
\end{equation}
Hence, the first of the equations \eqref{BD-BO} implies
\begin{equation}\label{BDnew}
	B_{D}=-\frac{\mathrm{i}}{2}Q^2\sigma_3.
\end{equation}
Also, the terms of order $1/k$ of $M_x \mu=0$ yield
\begin{equation}\label{order 1}
	\mu_{1x}+\mathrm{i}[\sigma_3,\mu_2]=Q\mu_1.
\end{equation}
Hence
\begin{equation}\label{consequence}
	\mathrm{i}[\sigma_3,\mu_{2O}]=Q\mu_{1D}-\mu_{1Ox}.
\end{equation}
Using this equation in the second of equations \eqref{BD-BO} we find
\begin{equation}\label{BOnew}
	B_O=-\frac{\mathrm{i}}{2}Q_x\sigma_3.
\end{equation}
Thus, if the jump matrix $J$ satisfies the conditions \eqref{commutations}, the RH problem \eqref{RH problem} is associated with the Lax pair
\begin{subequations}\label{lax pair}
	\begin{align}
		&\mu_x+\mathrm{i}k[\sigma_3,\mu]=Q\mu,\label{LPx}\\
		&\mu_t+\mathrm{i}k^2[\sigma_3,\mu]=\left[kQ-\frac{\mathrm{i}}{2}(Q_x+Q^2)\sigma_3\right]\mu, \qquad k\in\mathbbm{C}.\label{LPt}
	\end{align}
\end{subequations}
Letting
\begin{equation}
	\mu=\psi \displaystyle{e^{\mathrm{i}(kx+k^2t)\sigma_3}},\nonumber
\end{equation}
Eqs \eqref{lax pair} become
\begin{align}\label{lax pair psi}
	&\psi_x+\mathrm{i}k\sigma_3\psi=Q\psi, \nonumber \\ 
	&\psi_t+\mathrm{i}k^2\sigma_3\psi=\left[kQ-\frac{\mathrm{i}}{2}(Q_x+Q^2)\sigma_3\right]\psi, \qquad k\in\mathbbm{C}.
\end{align}
The compatibility condition of these equations is
\begin{equation}\label{NLS matrix}
	\mathrm{i}Q_t-\frac{1}{2}Q_{xx}\sigma_3+Q^3\sigma_3=0.
\end{equation}
Eq \eqref{Q} implies that $Q$ can be solved by computing the solution of \eqref{RH problem} and then identifying the off-diagonal part of the term of order $1/k$ (for large $k$) of the solution $\mu$.\\

In the particular case that $Q$ has the form
\begin{equation}
	Q=\begin{pmatrix} 0 &  q\cr \lambda \bar{q} & 0 \end{pmatrix},\qquad \lambda=\pm1,\label{Q matrix}
\end{equation} 
the \q{12} term of \eqref{NLS matrix} becomes equation \eqref{NLS}.\\

It should be noted that \eqref{NLS matrix} can also be derived by looking at the off diagonal terms of order $1/k$ (for large $k$) of \eqref{LPt}:
\begin{equation}\label{off diagonal}
	\mu_{1Ot}+\mathrm{i}[\sigma_3,\mu_{1O}]=Q\mu_{1D}-\frac{\mathrm{i}}{2}Q_x\sigma_3\mu_{1D}-\frac{\mathrm{i}}{2}Q^2\sigma_3\mu_{1O}.
\end{equation}
The terms of order $1/k^2$ of \eqref{LPx} yield
\begin{equation}\label{order 1 over k 2}
	\mu_{2Ox}+\mathrm{i}[\sigma_3,\mu_{2O}]=Q\mu_{2D}.
\end{equation}
Also, \eqref{consequence} yields
\begin{equation}\label{consequence order 2}
	\mu_{2O}=\frac{\mathrm{i}}{2}(Q\mu_{1D}-\mu_{1Ox})\sigma_3=\frac{\mathrm{i}}{2}Q\sigma_3\mu_{1D}+\frac{1}{4}Q_x,
\end{equation}
where we have used \eqref{mu1O}. Furthermore, the diagonal part of \eqref{order 1} yields
\begin{equation}\label{diagonal part}
	\mu_{1Dx}=Q\mu_{1O}=\frac{\mathrm{i}}{2}Q^2\sigma_3.
\end{equation}
Using \eqref{order 1 over k 2}-\eqref{diagonal part} in \eqref{off diagonal} we find \eqref{NLS matrix}.\\

We next show that  $\Delta_g$ does \textit{not} commute with $J$: let $\Delta_g\mu^{\pm}$ be defined by \eqref{pm} with $ f $ replaced by $ \mu $. Replacing in the expressions for $ \Delta_g \mu ^{-} $, $ \mu^{-} $ by $ \mu^{+}J $, we see that $ \Delta_g \mu ^{-} $ would simplify if we impose on $ J $ the evolution

\begin{equation}\label{J evolution}
	J_{t}+\frac{1}{2\mathrm{i}}(Hg)[\sigma , J]=0, \qquad k\in\mathbb{R}.
\end{equation}\\

However, even with these simplifications,  $ \Delta_g \mu ^{-} $ is not equal to  $ (\Delta_g \mu ^{+})J $:
\begin{equation}
	\Delta_g\mu^-=\left(\Delta_g\mu^+\right)J+\frac{1}{2\mathrm{i}}\left\{Hg[\mu^-\sigma_3(\mu^-)^{-1}-\mu^+\sigma_3(\mu^+)^{-1}]\right\}\mu^+J.
\end{equation}
In what follows, we will \textit{not} impose on $ J $ the evolution \eqref{J evolution} but we will determine the time-evolution by analysing directly the equation $ M_t\mu=0 $.\\

We define the new $ t $-part of the Lax pair by adding the term $ \Delta_g \mu$ to the old Lax pair
\begin{align}\label{general equation reduction}
	\mu^{\pm}_t+\mathrm{i}\alpha k^2[\sigma_3,\mu^{\pm}]-(Cg)\mu^{\pm}\sigma_3&+\left(Cg\mu^{\pm}\sigma_3(\mu^{\pm})^{-1}\right)\mu \notag\\
	&=\alpha\left[kQ-\frac{\mathrm{i}}{2}(Q_x+Q^2)\sigma_3\right]\mu^{\pm},\quad \text{Im}k\gtrless 0.\TAG{$ (48)^{\pm} $}
\end{align}\refstepcounter{equation}
We have introduced the constant parameter $ \alpha $ in order to be able to consider the novel integrable PDE arising in the limit of $ \alpha=0 $.\\

The terms in \ref{general equation reduction} $ \hspace{1pt} $ involving $ g $ vanish for large $ k $, thus the terms of $ O(k) $ and $ O(1) $, as $ k \rightarrow \infty $, of the new $ t $-part of the Lax pair are identical to the corresponding terms of the old $ t $-part; thus these terms vanish. We expect that $ M_t\mu =0$ will be compatible with  $ M_x\mu =0$ provided that the terms of order $ 1/k $, for large $ k $, of $ M_t\mu $ vanish. In this connection we note that the off-diagonal part of these terms are

\begin{equation}\label{A_10}
	\frac{1}{2\mathrm{i}\pi}\int_\mathbb{R}g\left(\mu\sigma_3\mu^{-1}\right)_O  \hskip1mm dl. \nonumber
\end{equation}\\

In summary, let $J$ satisfy \eqref{commutations}.  Then, the RH problem    \eqref{RH problem} is associated with the equation \eqref{LPx}.  Furthermore, if $\mu$ satisfies \ref{general equation reduction}$ \hspace{3pt} $, this equation is compatible with the equation \eqref{LPx}. The off-diagonal term of the order $1/k$ (for large $k$) of the solution of the RH problem \eqref{RH problem} satisfies the equation
\begin{equation}\label{solution of RH}
	\mathrm{i}Q_t-\frac{\alpha}{2}Q_{xx}\sigma_3+\alpha Q^3\sigma_3=-\frac{1}{\mathrm{i}\pi}\int_\mathbb{R}g(t,l)\left(\mu\sigma_3\mu^{-1}\right)_O(x,t,l)  \hskip1mm  dl  \hskip1mm \sigma_3.
\end{equation}
In the particular case of $\alpha=0$, we find the following set of equations:
\begin{equation}\label{x and t evolution for J}
	J_x+\mathrm{i}k[\sigma_3,J]=0, \qquad J_t+\frac{1}{2\mathrm{i}}(Hg)[\sigma_3,J]=0,\qquad k\in \mathbb{ R},
\end{equation}
\begin{equation}\label{x and t evolution for mu}
	\mu_x+\mathrm{i}k[\sigma_3,\mu]-Q\mu=0, \qquad \mu_t=(Cg)\mu\sigma_3-\left(Cg\mu\sigma_3\mu^{-1}\right)\mu,
\end{equation}
\begin{equation}\label{truncated equation}
	\mathrm{i}Q_t=-\frac{1}{\mathrm{i}\pi}\int_\mathbb{R}g(t,l)\left(\mu\sigma_3\mu^{-1}\right)_O (x,t,l) \hskip1mm  dl  \hskip1mm \sigma_3.
\end{equation}\\

\vskip5mm

\noindent \underline{Method of Solution}\\

We define the operator $\displaystyle{\widehat \sigma}$ by
\begin{equation}\label{sigma hat}
	{\widehat \sigma}f=[\sigma,f].
\end{equation}
This definition implies that 
\begin{equation}\label{matrix}
	\mathrm{e}^{\gamma{\widehat \sigma_3}}f=\mathrm{e}^{\gamma\sigma_3}f\mathrm{e}^{-\gamma\sigma_3}=\begin{pmatrix} f_{11} &  \mathrm{e}^{2\gamma} f_{21}\cr\mathrm{e}^{-2\gamma} f_{12} & f_{22} \end{pmatrix},
\end{equation}
where $\gamma$ is an arbitrary scalar and $f_{ij}$ are the entries of the matrix $f$.\\

It is important to note that $ \Delta_g\mu^{+} $ and $ \Delta_g\mu^{-} $ satisfy the \textit{same} equations. Thus, if $X (x, t, k)$ denotes either $ \mu^{+} $ or $ \mu^{-} $ for $ k $ real, it follows that $ X $ satisfies the Lax pair \eqref{x evolution of X}-\eqref{t evolution of X}. Similarly, $ \mu $ satisfies a similar Lax pair when $ H/2\mathrm{i} $ is replaced by the Cauchy operator $ C $.\\

We define the following particular solutions of \eqref{x evolution of X}:
\begin{equation}\label{phi}
	\Phi(x,t,k)=I+\int_{-\infty}^x d\xi\mathrm{e}^{\mathrm{i}k(\xi -x){\widehat \sigma_3}}Q(\xi,t)\Phi(\xi,t,k),
\end{equation}
and
\begin{equation}\label{psi}
	\Psi(x,t,k)=I-\int_{x}^\infty d\xi\mathrm{e}^{\mathrm{i}k(\xi -x){\widehat \sigma_3}}Q(\xi,t)\Psi(\xi,t,k).
\end{equation}
Using the second equality of \eqref{matrix} with $\gamma=2\mathrm{i}k(\xi-x)$, it follows that
\begin{equation}\label{phi-psi}
	\Phi=(\Phi^+,\Phi^-),\qquad \Psi=(\Psi^-,\Psi^+),
\end{equation}
where $\Phi^+$ and  $\Phi^-$ are the first and second column vectors of the matrix  $\Phi$, and the notation $+$, $-$ denotes analyticity in $k$ for $\text{Im}k >0$,  $\text{Im}k <0$, respectively. Indeed, in the integral appearing in \eqref{phi}, $x- \xi$ is non-negative, hence $\exp [ 2\mathrm{i}k( x- \xi)]$ is bounded and analytic for $\text{Im}k >0$. Assuming that $Q$ has sufficient smoothness and decay, Eq \eqref{phi} is a Volterra integral equation and hence the $k$ dependence of the solution inherits the properties of  $\exp [ 2\mathrm{i}k( x- \xi)]$. Similar considerations are valid for $\Psi$.\\

Eq \eqref{x evolution of X} implies that $X$ satisfies $(\text{det} X) _ x=0$, hence $\text{det} \Phi=\text{det} \Psi=1$. Furthermore, if $Q$ is given by \eqref{Q matrix}, then $\Phi$ and $\Psi$ satisfy the following symmetry relations:
\begin{align}\label{matrix elements}
	&{\overline {\Phi_1^+(k)}}={ {\Phi_2^-(\overline {k})}},\quad
	{\overline {\Phi_2^+(k)}}=\lambda{ {\Phi_1^-(\overline {k})}},\nonumber \\
	&{\overline {\Psi_1^-(k)}}={ {\Psi_2^+(\overline {k})}},\quad
	{\overline {\Psi_2^-( {k})}}=\lambda\Psi_1^+(\overline {k}),
\end{align}
where for convenience we have suppressed the $x$ and $t$ dependence.\\
Since both $\Phi$ and $\Psi$ satisfy \eqref{x evolution of X}, they are  related by an $x$-independent matrix $S(t,k)$:
\begin{equation}\label{phi psi related}
	\Phi(x,t,k)=\Psi(x,t,k)\mathrm{e}^{-\left(\mathrm{i}kx\right)\widehat{\sigma}_3}S(t,k),\qquad k\in\mathbbm{R}.
\end{equation}
The operator $\exp(\gamma \widehat\sigma_3)$ satisfies the identity
\begin{equation}\label{exp relation}
	\mathrm{e}^{\gamma \widehat\sigma_3}FG=\left(  \mathrm{e}^{\gamma \widehat\sigma_3} F\right) \mathrm{e}^{\gamma \widehat\sigma_3} G.\nonumber
\end{equation}
Multiplying \eqref{phi psi related} by $\mathrm{e}^{-\left(\mathrm{i}kx\right)\widehat{\sigma}_3}$ and evaluating the resulting equation as $x$ tends to $-\infty$, we find
\begin{equation}\label{limit}
	S(t,k)=\lim_{x\rightarrow -\infty}\left(\mathrm{e}^{\mathrm{i}kx\widehat{\sigma}_3}\Psi(x,t,k)\right)^{-1}.
\end{equation}
The symmetry relations \eqref{matrix elements} imply that
\begin{align}\label{new S matrix}
	S(t,k)=\begin{pmatrix}  \overline{a}(t,k) &  b(t,k)\cr \lambda \overline{b}(t,k) &{a}(t,k) \end{pmatrix},\qquad k\in \mathbb{R}, \qquad t>0.
\end{align}

The function $ \Psi(x,t,k) $ satisfies eq \eqref{voltera}. Multiplying this equation by $ \exp(\mathrm{i}kx\widehat{\sigma}_3) $, letting $ x\rightarrow -\infty $, and using the definition \eqref{limit} we find
\begin{align}\label{Sevolution}
	S_t^{-1}+\mathrm{i}\alpha k^{2}\widehat{\sigma}_3S^{-1}&=\frac{1}{2\mathrm{i}}(Hg)S^{-1}\sigma_3
	\nonumber \\
	&-\frac{1}{2\mathrm{i}}\lim_{x\rightarrow -\infty} \left\{\mathrm{e}^{\mathrm{i}kx\widehat{\sigma}_3}\left(Hg\mathrm{e}^{-\mathrm{i}kx{\sigma_3}}S^{-1}\sigma_3S\mathrm{e}^{\mathrm{i}kx{\sigma_3}}\right)\right\}S^{-1}, \quad k\in \mathbb{R}.
\end{align}
For the simplification of this equation we have used the following identities:
\begin{align}\nonumber
	&	\left(   \mathrm{e}^{-\mathrm{i}kx{\widehat\sigma_3}}S^{-1}   \right)   \sigma_3
	\left(     \mathrm{e}^{-\mathrm{i}kx{\widehat\sigma_3}}    S\right)=
	\mathrm{e}^{-\mathrm{i}kx{\widehat\sigma_3}}S^{-1}\sigma_3 S
	\mathrm{e}^{\mathrm{i}kx{\widehat\sigma_3}},	
	\nonumber \\
	&\mathrm{e}^{\mathrm{i}kx{\widehat\sigma_3}}
	\left\{ F\left(  \mathrm{e}^{-\mathrm{i}kx{\widehat\sigma_3}}S^{-1}  \right) \right\}=
	\left\{\mathrm{e}^{\mathrm{i}kx{\widehat\sigma_3}}F	\right\}S^{-1}.
	\nonumber
\end{align}
Using the definition of $ S $, we find
\begin{equation*}
	\mathrm{e}^{-\mathrm{i}kx\sigma_3}S^{-1}\sigma_3S\mathrm{e}^{\mathrm{i}kx\sigma_3}=	
	\begin{pmatrix}
		|a|^{2}+\lambda|b|^{2}	&  2ab\mathrm{e}^{-2\mathrm{i}kx}\\
		-2\lambda\bar{a}\bar{b}\mathrm{e}^{2\mathrm{i}kx}	& -\left(|a|^{2}+\lambda|b|^{2}\right)
	\end{pmatrix}	.
\end{equation*}
Thus,
\begin{equation*}
	\lim_{x \rightarrow -\infty}\left(     Hg\mathrm{e}^{-\mathrm{i}kx{\sigma_3}}S^{-1}\sigma_3S\mathrm{e}^{\mathrm{i}kx{\sigma_3}}           \right)=
	\begin{pmatrix}
		Hg\left( |a|^{2}+\lambda|b|^{2}  \right)	& 0 \\
		0	&-\left(   Hg\left( |a|^{2}+\lambda|b|^{2}  \right)   \right)
	\end{pmatrix},	
\end{equation*}
where we have used the fact that, for any sufficiently decaying function $ f $,
\begin{equation*}
	\lim_{x\rightarrow-\infty}\left(   Hf\mathrm{e}^{\pm 2\mathrm{i}kx}\right)=0.
\end{equation*}
Thus, \eqref{Sevolution} becomes
\begin{equation}\label{Sevolution simplified}
	S_t^{-1}+\mathrm{i}\alpha k^{2}\widehat{\sigma}_3S^{-1}=\frac{1}{2\mathrm{i}}(Hg)S^{-1}\sigma_3
	-\frac{1}{2\mathrm{i}}\left( Hg \left(      |a|^{2}+\lambda|b|^{2}         \right)         \right)\sigma_3S^{-1}, \quad k\in \mathbb{R}.
\end{equation}
This equation together with the fact that $ \text{det}(S)=1 $, namely
\begin{equation*}
	|a|^{2}-\lambda|b|^{2}=1,
\end{equation*}
yield the following evolution equations for $ a $ and $ b $:
\begin{subequations}\label{2 elements}
	\begin{align}
		&{a}_t=\lambda\mathrm{i}\left(Hg|b|^2\right){a},\label{2 elements 1}\\
		&b_t-2\mathrm{i}\alpha k^2b=\mathrm{i}\left[(Hg)+\lambda\left(Hg|b|^2\right)\right]b=0.\label{2 elements 2}
	\end{align}
\end{subequations}
These equations imply that $|a|$ and $|b|$ are $t$-independent. Thus,
\begin{align}\label{intermediary relations}
	&a(t,k)=a_0(k)\mathrm{e}^{\mathrm{i}\lambda \left(HG(t,k)|b_0|^2(k)\right)}, \nonumber\\ 
	&b(t,k)=b_0(k)\mathrm{e}^{2\mathrm{i}\alpha k^2t+\mathrm{i}\left(HG(t,k)\right)+\mathrm{i}\lambda\left( HG(t,k)|b_0|^2(k)\right)},\qquad k\in \mathbb{R}.
\end{align}

Writing in equation \eqref{phi psi related}, $\Phi$ and $\Psi$ in the form \eqref{phi-psi} and rearranging we find that \eqref{phi psi related} takes the form \eqref{mu matriciel}, where
\begin{equation}\label{new scattering relations}
	\frac{b(t,k)}{\overline{a}(t,k)}=\frac{b_0(t)}{\overline{a}_0(t)}\mathrm{e}^{2\mathrm{i}\alpha k^2t+\mathrm{i}\left(HG(t,k)\right)}.\nonumber
\end{equation}
In the particular case that $g$ is independent of $t$, the above equation becomes

\begin{equation}
	\frac{b}{\overline{a}}(t,k)=\frac{b_0(k)}{\overline{a}_0(k)}\mathrm{e}^{2\mathrm{i}\alpha k^2t+\mathrm{i}\left(Hg(k)\right)t},\nonumber
\end{equation}
where the matrices $\mu^{+}$, $\mu^{-}$  are defined as follows:
\begin{align}\label{matrices mu +-}
	&\mu^-(x,t,k)=\left(\Psi^-(x,t,k), \frac{\Phi^-(x,t,k)}{\overline{a(\overline{k},t)}}\right),
	\nonumber \\
	&\mu^+(x,t,k)=\left(\frac{\Phi^+(x,t,k)}{a(k,t)},\Psi^+(x,t,k) \right).
\end{align}\\

\noindent \underline{An analycity constraint on $ g(t,k) $}\\

In order for $ a(t,k) $ to be analytic for $ \text{Im}k>0 $ it is necessary for the function $ HG|b_0|^{2} $ to admit an analytic continuation in the upper half $ k $-complex plane. In this connection we note that 
\begin{equation*}
	\mathrm{i}HG|b_0|^{2}  =\left(\mathrm{i} HG|b_0|^{2}  -G|b_0|^{2} \right)+HG|b_0|^{2},         
\end{equation*}
and the first term in the RHS of the above equation admits an analytic continuation in $ \text{Im}k>0 $. Thus, we require that
\begin{equation*}
	G(k,t)|b_0(k)|^{2}=\tilde{g}(t,k), \qquad k\in \mathbb{R},
\end{equation*}
where $ \tilde{g}(t,k) $ is analytic in  $ \text{Im}k>0 $.\\

\noindent \underline{Rigorous considerations}\\

The above construction made explicit use of the matrices $Q$ and $Q _x$ which are unknown, hence this construction is purely formal. However, it can be rigorously justified a posteriori   as follows: given a function $q_0(x)$ in $L_1$, the first of the Volterra integral equations \eqref{voltera} yields the unique vector $\Psi^+_0(x, k)$. Then, Eqs \eqref{intermediary relations} define the scalars $a_0(k)$ and $b_0(k)$. Since $a(k,t)$ is analytic for $\text{Im}k >0$, the functions  $\mu^+$  and $\mu^- $ are analytic  for  $\text{Im}k >0$,  and   $\text{Im}k <0$ respectively, provided that $ a_0(k)$ does not have any zeros for  $\text{Im}k >0$. We assume that this is indeed the case. The occurrence of zeros corresponds to solitons, which, as discussed in appendix B, can be handled in an algebraic manner \cite{doi:10.1137/0527040}. Eq \eqref{mu matriciel} supplemented with \eqref{RH problem1} define a RH problem. 
If $ q_0\in L_1\cap L_2 $ and if the Hilbert transform of $ g $ has sufficient decay, then this RH problem is equivalent to a linear Fredholm integral equation.
The jump matrix $J$ has unit determinant, and also the complex conjugate of its \q{12} element equals $+$ or $-$ the value \q{21}. This symmetry property implies a \q{vanishing lemma}, namely, this RH problem has a unique solution \cite{Fokas_2005}. We define $Q$ in terms of this solution via \eqref{Q}. We need to prove that the \q{12} element of $ q $ solves eq. \eqref{full NLS} and thus it satisfies the initial conditions.

The \q{12} element of $[\mu^+\sigma_3(\mu^+)^{-1}]\sigma_3$ is the \q{12} element of the matrix
\begin{equation}\label{last nonumber 3}
	\begin{pmatrix} \frac{\Phi_1^+}{a} &  \Psi^+_1\cr \frac{\Phi_2^+}{a} &\Psi^+_2 \end{pmatrix}\sigma_3
	\begin{pmatrix} \Psi^+_2& -\Psi^+_1\cr  -\frac{\Phi_2^+}{a} & \frac{\Phi_1^+}{a} \end{pmatrix}
	\sigma_3.
\end{equation}
This yields $2\Psi_1^+\Phi^+_1/{a}$, and then the RHS of \eqref{full NLS} follows. The fact that the above $ q $ satisfies \eqref{full NLS} is a consequence of the DLT, which is discussed in the next section. \\

The above construction only used the unique solvability of the RH problem \eqref{RH problem}. The proof that $q$ evaluated at $t=0$ coincides with the value of $q_0$, is a consequence of the construction of an auxiliary RH problem characterizing $q_0$ via $a_0(k)$ and $b_0(k)$ \cite{Fokas_2005}.\\

It should be noted that the matrices $\mu^+$ and $\mu^- $  must satisfy both the $x$ and $ t$ parts of the Lax pair. Since $a_0(k)$ is independent of $x$ and $t$, and the vectors forming $\Phi$ and $\Psi$ satisfy the Lax pair, the matrices $\mu^+$ and $\mu^-$   defined by (3.14) also satisfy the same Lax pair. 

\section{The Direct Linearizing Transform (DLT)}

Before implementing the DLT it is worth noting that for linear PDEs, the main weakness of the Ehrenpreis result is that it does not provide a method for computing the measure. This is to be contrasted with the unified transform \cite{10.2307/53053} which computes the measure $d\rho(l)$ explicitly in terms of the initial and boundary conditions. Similarly, the direct linearizing transform, in contrast to the inverse scattering and unified transforms, does not relate the measure to initial and boundary conditions. However, as shown by many authors, and in particular by Frank Nijhof \cite{NIJHOFF1988339}\cite{Nijhoff_1990}, it does provide a powerful approach for computing large classes of solutions of the associated PDEs. Also, it should be noted that the assumption that the domain is smooth and bounded is not needed (but convexity is essential). Indeed, the unified transform, which always yields representations in the Ehrenpreis form, has been implemented to domains which are unbounded and have corners. \\

In what follows we derive the direct linearizing transform for the nonlinear integrable equation \eqref{truncated equation} associated with the Lax pair \eqref{x and t evolution for mu}.  After all, the terms associated with the NLS equation, in contrast to the novel terms involving the Hilbert transform, have been considered before. Also, in order to simplify the presentation of the proof, we assume that $g$ is $t$-independent.\\

In order to motivate the formulation of the direct linearizing transform we start with the RH problem with the jump condition \eqref{mu matriciel}, which we rewrite in the following form:
\begin{equation}\label{new jump function}
	\mu^-(x,t,k)-\mu^+(x,t,k)=\mu^-(x,t,k)\displaystyle{\mathrm{e}^{\nu(x,t,k){\widehat\sigma_3} }}\widetilde{J}(k),\qquad k\in \mathbb{R},
\end{equation}
where the function $\nu$ and the new jump matrix, $\widetilde{J}(k)$, are defined   by
\begin{align}\label{jump J}
	&\nu(x,t,k)=-\mathrm{i}kx-\mathrm{i}\alpha k^2t+\frac{\mathrm{i}}{2}(Hg)(k)t,\nonumber \\
	&\widetilde{J}(k)=\begin{pmatrix} 0 &  -(\frac{b_0(k)}{\overline{a_0}(k)}+1)\cr \lambda \frac{\overline{b_0}(k)}{a_0(k)}-1 & \frac{1}{|a_0(k)|^2}-1 \end{pmatrix},\qquad k\in \mathbb{R}.
\end{align}
Equation \eqref{new jump function} together with the assumption that $\mu$ tends to the identity for large $k$, imply
\begin{equation}\label{integral mu}
	\mu(x,t,k)=I+\int_{\mathbb{ R}}\frac{dl}{l-k}\mu^-(x,t,l)\mathrm{e}^{\nu(x,t,l)\widehat{\sigma}_3}\widetilde{J}(l),\qquad k\in \mathbb{C}.
\end{equation}
By taking the limit of $k$ as it approaches the real axis from below, we obtain a linear integral equation for $\mu$. This motivates the following result:\\

\def\@thmcountersep{.}
\newtheorem*{Prop}{Proposition 1}
\newtheorem*{Dem}{Proof}
\begin{Prop}\label{proposition}
	Let $X (x, t, k)$ satisfy the linear integral equation
	\begin{equation}\label{X_integral equation}
		X(x,t,k)=I+\int_L\frac{d\rho(l)}{l-k}X(x,t,l)\mathrm{e}^{\nu(x,t,l)\widehat{\sigma}_3}\widetilde{J}(l),\qquad k\in \mathbb{C},
	\end{equation}
\end{Prop}
\noindent where the measure $ d\rho(l) $ and the contour $ L $ are such that the RHS of \eqref{X_integral equation} is well defined. For example, $ d\rho(l) =dl/2\pi\mathrm{i}$ and $ L $ the real axis, $ X=X^{-} $ and $ k $ is restricted to the lower $ k $-complex plane. Assume that the homogenous version of \eqref{X_integral equation}, namely the equation obtained from \eqref{X_integral equation} by replacing $ I $ with zero, has only the zero solution. Then, $X$ satisfies the Lax pair \eqref{x and t evolution for mu} where $ Q(x,t) $ is defined by
\begin{equation}\label{eq10}
	Q(x,t)=\mathrm{i}\widehat{\sigma}_3\int_{ L}d\rho(l)X(x,t,l)\mathrm{e}^{\nu(x,t,l)\widehat{\sigma}_3}\widetilde{J}(l).
\end{equation}

\begin{Dem}
	We first consider the $x$-part of the Lax pair. Differentiating \eqref{X_integral equation} with respect to $x$ we find
	\begin{equation}\label{Derivative of X}
		X_x=\int_L\frac{d\rho(l)}{l-k}X_x\mathrm{e}^{\nu\widehat{\sigma}_3}\widetilde{J}+\int_L\frac{d\rho(l)}{l-k}\mathrm{i}lX\mathrm{e}^{\nu\widehat{\sigma}_3}(\widehat{\sigma}_3\widetilde{J}).
	\end{equation}
	Applying the operator $\widehat{\sigma}_3$  to \eqref{X_integral equation} we find
	\begin{equation}\label{sigma 3 Derivative of X}
		\widehat{\sigma}_3X=\int_L\frac{d\rho(l)}{l-k}(\widehat{\sigma}_3X)\mathrm{e}^{\nu\widehat{\sigma}_3}\widetilde{J}+\int_L\frac{d\rho(l)}{l-k}X\mathrm{e}^{\nu\widehat{\sigma}_3}(\widehat{\sigma}_3\widetilde{J}),
	\end{equation}
	where we used the identity
	\begin{equation}\label{idendity}
		\widehat{\sigma}_3\left(X\mathrm{e}^{\nu\widehat{\sigma}_3}\widetilde{J}\right)=(\widehat{\sigma}_3X)\mathrm{e}^{\nu\widehat{\sigma}_3}\widetilde{J}+X\mathrm{e}^{\nu\widehat{\sigma}_3}(\widehat{\sigma}_3\widetilde{J}).
	\end{equation}
	Multiplying Eq \eqref{sigma 3 Derivative of X} by $\mathrm{i}k$ and then adding and subtracting the term
	\begin{equation}\label{unnamed 1}
		\int_L\mathrm{i}l\frac{d\rho(l)}{l-k}(\widehat{\sigma}_3X)\mathrm{e}^{\nu\widehat{\sigma}_3}\widetilde{J},\nonumber
	\end{equation}
	we find 
	\begin{equation}\label{sigma3X expanded}
		\widehat{\sigma}_3X=\int_L\frac{d\rho(l)}{l-k}\mathrm{i}l(\widehat{\sigma}_3X)\mathrm{e}^{\nu\widehat{\sigma}_3}\widetilde{J}+\int_L\frac{d\rho(l)}{l-k}\mathrm{i}kX\mathrm{e}^{\nu\widehat{\sigma}_3}(\widehat{\sigma}_3\widetilde{J})+\mathrm{i}\int_{ L}d\rho(l)(\widehat{\sigma}_3X)\mathrm{e}^{\nu\widehat{\sigma}_3}\widetilde{J}.
	\end{equation}
	Multiplying \eqref{X_integral equation} by $-Q$ we find
	\begin{equation}\label{QX}
		-QX=-Q+\int_L\frac{d\rho(l)}{l-k}(-Q)X\mathrm{e}^{\nu\widehat{\sigma}_3}\widetilde{J}.
	\end{equation}
	Adding \eqref{Derivative of X}, \eqref{sigma3X expanded}, and \eqref{QX} we find that the function
	\begin{equation}\label{unnamed 2}
		X_x+\mathrm{i}k\widehat{\sigma}_3X-QX \nonumber
	\end{equation}
	satisfies the homogeneous version of \eqref{X_integral equation} provided that $Q(x,t)$ satisfies
	\eqref{eq10}, 
	where we have used again the identity \eqref{idendity}.\\
	
	We next consider the $t$-part of the Lax pair. Differentiating \eqref{X_integral equation} with respect to $t$ we find 
	\begin{equation}\label{time evolution of mu}
		X_t=\int_L\frac{d\rho(l)}{l-k}X_t\mathrm{e}^{\nu\widehat{\sigma}_3}\widetilde{J}+\int_L\frac{d\rho(l)}{l-k}\frac{\mathrm{i}}{2}(Hg)(l)X\mathrm{e}^{\nu\widehat{\sigma}_3}({\widehat{\sigma}_3}\widetilde{J}).
	\end{equation}
	Multiplying \eqref{X_integral equation} from the right by $\sigma_3$ and then adding and subtracting the term
	\begin{equation}\label{unnamed 3}
		\int_L\frac{d\rho(l)}{l-k}X\sigma_3\mathrm{e}^{\nu\widehat{\sigma}_3}\widetilde{J}, \nonumber
	\end{equation}
	we find 
	\begin{equation}\label{mu sigma3}
		X\sigma_3=\sigma_3+\int_L\frac{d\rho(l)}{l-k}X\sigma_3\mathrm{e}^{\nu\widehat{\sigma}_3}\widetilde{J}-\int_L\frac{d\rho(l)}{l-k}X\mathrm{e}^{\nu\widehat{\sigma}_3}(\widehat{\sigma}_3\widetilde{J}).
	\end{equation}
	Multiplying this equation by $\mathrm{i}(Hg)(k)/2$ and then adding and subtracting the term
	\begin{equation}\label{unnamed 4}
		\int_L\frac{d\rho(l)}{l-k}\frac{\mathrm{i}}{2}(Hg)(l)X\sigma_3\mathrm{e}^{\nu\widehat{\sigma}_3}\widetilde{J}, \nonumber
	\end{equation}
	we find
	\begin{align}\label{we find}
		\frac{\mathrm{i}}{2}&(Hg)(k)X\sigma_3=\frac{\mathrm{i}}{2}(Hg)(k)\sigma_3+\int_L\frac{d\rho(l)}{l-k}\frac{\mathrm{i}}{2}(Hg)(l)X\sigma_3\mathrm{e}^{\nu\widehat{\sigma}_3}\widetilde{J} \nonumber\\
		&+\int_L\frac{d\rho(l)}{l-k}\frac{\mathrm{i}}{2}\left[(Hg)(k)-(Hg)(l)\right]X\sigma_3\mathrm{e}^{\nu\widehat{\sigma}_3}\widetilde{J}-\int_L\frac{d\rho(l)}{l-k}\frac{\mathrm{i}}{2}(Hg)(k)X\mathrm{e}^{\nu\widehat{\sigma}_3}({\widehat{\sigma}_3}\widetilde{J}).
	\end{align}
	Multiplying \eqref{X_integral equation} by $-\mathrm{i}F/2$, where $F$ is defined by 
	\begin{equation}\label{definition of F}
		F(x,t,k)=(HgX\sigma_3X^{-1})(x,t,k),
	\end{equation}
	we find
	\begin{align}\label{F times mu}
		-\frac{\mathrm{i}}{2}FX=-\frac{\mathrm{i}}{2}F-&\frac{\mathrm{i}}{2}\int_L\frac{d\rho(l)}{l-k}FX\mathrm{e}^{\nu\widehat{\sigma}_3}\widetilde{J}
		\nonumber \\
		&-\frac{\mathrm{i}}{2}\int_L\frac{d\rho(l)}{l-k}\big[F(x,t,k)
		-F(x,t,l)\big]X\mathrm{e}^{\nu\widehat{\sigma}_3}\widetilde{J}.
	\end{align}
	Adding \eqref{time evolution of mu}, \eqref{we find}, and \eqref{F times mu} we find that the function defined by the second of the equations \eqref{x and t evolution for mu} satisfies the homogeneous version of \eqref{X_integral equation} provided the  following equation is valid:
	\begin{align}
		\frac{\mathrm{i}}{2}(Hg)(k)\sigma_3-\frac{\mathrm{i}}{2}F(x,t,k)&-\frac{\mathrm{i}}{2}\int_L\frac{d\rho(l)}{l-k}\left[F(x,t,k)-F(x,t,l)\right]X\mathrm{e}^{\nu\widehat{\sigma}_3}\widetilde{J} \nonumber \\
		&+\frac{\mathrm{i}}{2}\int_L\frac{d\rho(l)}{l-k}\left[(Hg)(k)-(Hg)(l)\right]X\mathrm{e}^{\nu\widehat{\sigma}_3}\widetilde{J}\sigma_3=0.
		\nonumber
	\end{align}
	Simplifying this equation, we find
	\def\Xint#1{\mathchoice
		{\XXint\displaystyle\textstyle{#1}}%
		{\XXint\textstyle\scriptstyle{#1}}%
		{\XXint\scriptstyle\scriptscriptstyle{#1}}%
		{\XXint\scriptscriptstyle\scriptscriptstyle{#1}}%
		\!\int}
	\def\XXint#1#2#3{{\setbox0=\hbox{$#1{#2#3}{\int}$ }
			\vcenter{\hbox{$#2#3$ }}\kern-.6\wd0}}
	\def\ddashint{\Xint=}
	\def\dashint{\Xint-}
	\begin{align}\label{almost done}
		(Hg)(k)\sigma_3-F(x,t,k)&+\frac{P}{\pi}\int dl\rq{}\int_L\frac{d\rho(l)g(l\rq{})(X\sigma_3X^{-1})(l\rq{})}{(l\rq{}-k)(l\rq{}-l)}X\mathrm{e}^{\nu\widehat{\sigma}_3}\widetilde{J} \nonumber \\
		&-\frac{P}{\pi}\int dl\rq{}\int_L\frac{d\rho(l)}{(l\rq{}-k)(l\rq{}-l)}g(l\rq{})X\mathrm{e}^{\nu\widehat{\sigma}_3}\widetilde{J}\sigma_3=0,
	\end{align}
	where  for convenience of notation we have suppressed the $x$ and $t$ dependence of $X$.
	Multiplying from the left the equation
	\begin{equation}\label{final}
		X(l\rq{})-I+\int_L\frac{d\rho(l)}{(l\rq{}-l)}X\mathrm{e}^{\nu\widehat{\sigma}_3}\widetilde{J}=0,
	\end{equation}
	by
	\begin{equation}\label{unnamed 6}
		\frac{1}{\pi}\frac{g(l\rq{})(X\sigma_3X^{-1})(l\rq{})}{l\rq{}-k},\nonumber
	\end{equation}
	and integrating over $d\rho(l\rq{})$ we find
	\begin{align}\label{unnamed 7}
		\frac{P}{\pi}\int \frac{dl\rq{}}{l\rq{}-k}g(l\rq{})X(l\rq{})\sigma_3&-F(x,t,k)
		\nonumber \\
		&+\frac{P}{\pi}\int dl\rq{}\int_L\frac{d\rho(l)g(l\rq{})(X\sigma_3X^{-1})(l\rq{})}{(l\rq{}-k)(l\rq{}-l)}X\mathrm{e}^{\nu\widehat{\sigma}_3}\widetilde{J}=0.\nonumber
	\end{align}
	Using this equation in \eqref{almost done} we obtain
	\begin{equation}
		(Hg)(k)\sigma_3-\frac{P}{\pi}\int\frac{dl\rq{}}{l\rq{}-k}g(l\rq{})X(l\rq{})\sigma_3-\frac{P}{\pi}\int dl\rq{}\int_L\frac{d\rho(l)g(l\rq{})}{(l\rq{}-k)(l\rq{}-l)}X\mathrm{e}^{\nu\widehat{\sigma}_3}\widetilde{J}\sigma_3=0.
		\nonumber
	\end{equation}
	Multiplying this equation from the right by $-\sigma_3$ , it becomes evident that the resulting equation can be obtained by multiplying \eqref{final} by $g(l\rq{})/(l\rq{} -k)$, and then integrating over $l\rq{}$.    
\end{Dem}
\noindent QED\\

The RH problem \eqref{RH problem} with the jump matrix $ \mathrm{e}^{\nu{\widehat \sigma_3}}\tilde{J} $ is equivalent to the linear integral equation \eqref{X_integral equation} when $ d\rho(l) =dl/2\mathrm{i}\pi$ and $ L $ is the real axis. Thus, if $ Q $ is defined by \eqref{eq10} which is equivalent to \eqref{Q}, then this $ Q $ satisfies the Lax pair \eqref{x and t evolution for mu}. Moreover, since the forced nonlinear PDE \eqref{full NLS} is obtained by the $ 1/k $ terms (for large $ k $) of $ M_t\mu=0 $, this $ Q $ solves \eqref{full NLS}.

\section{Physical applications}

In this section, we discuss two \textit{illustrative} examples of important physical phenomena associated with a forced version of integrable nonlinear evolution equations, where the forcing consists of terms involving certain eigenfunctions of the associated Lax pair. \\

The first example is the so-called  self-induced transparency phenomenon. This example is chosen in  order to illustrate the particular case of $ \alpha=0 $ in \eqref{full NLS} where both nonlinearity and dispersion are absent and the phenomenon is governed by the \q{forcing} term in the RHS. We expect that many other \q{forced} integrable systems governed only by forcing are also physically significant.\\

The second example is the Stimulated Raman Backscattering.  To our knowledge, the mapping presented in our work, which provides an explicit relation between the equations describing this physical situation and our mathematical formalism, is new and it establishes the integrability of this important physical system. In particular, our analysis reveals the specific ratio of the plasma to laser frequency, for which the Stimulated Raman Scattering is integrable. The existence of such a value was predicted in \cite{SRBS_Zakharov_Manakov}.\\

\noindent{\underline{Self-induced transparency}}\\

Consider the propagation of a laser pulse in a dielectric medium. When the frequency of the input beam is close to the transition frequency, then the dielectric is modelled as a two level medium and the incident laser pulse is expected to be strongly absorbed. However, it has been discovered 
\cite{PhysRev.183.457} that beyond some intensity threshold, for laser pulses much shorter than the relaxation time of the two-level medium, the later becomes almost transparent and the laser beam propagates with a very low energy loss. This phenomenon can be explained as follows:  A part of the input pulse\rq{}s energy excites the atoms from the ground state into the upper layer,  whilst the other part of it\rq{}s energy stimulates the atoms making them return to the ground state. In this process, atoms emit coherently  the initially absorbed energy. This coherent emission is added to the rear of the non absorbed pump in such a way that the net result is an electromagnetic pulse propagating without any energy loss. Hence, the material appears to be transparent to the input pulse. This phenomenon as a permanent exchange of energy between the input pulse and the medium can be described as follows:\\

In the slowly varying envelope approximation, the Maxwell-Bloch system is expressed in its dimensionless form  \cite{Foot:1080846} by the equations:
\begin{subequations}\label{Self induced transparency}
	\begin{align}
		&\mathcal{E}_\xi(\xi,\tau)=\int_{-\infty}^{+\infty}dl \hskip1mm g(l)f(\xi,\tau,l), \label{Self induced transparency 1 }\\
		&f_{\tau}(\xi,\tau,k)+2\mathrm{i}kf(\xi,\tau,k)=\mathcal{E}N(\xi,\tau,k),\\
		&N_{\tau}=-\frac{1}{2}(\overline{\mathcal{E}}f+\mathcal{E}\overline{f}),
	\end{align}
\end{subequations}
where $\mathcal{E}$ is the envelope of the electric field in the rotating frame $(\xi,\tau)$ co-moving with the incident  pulse, $f$ is the polarization ($\text{Re}(f)$ and $\text{Im}(f)$ are the in-phase and the quadrature components respectively), and $N$ is the population inversion. The parameter $k$ measures the difference between the frequency of the input pulse $\mathcal{E}$ and the resonant frequency gap. The RHS of \eqref{Self induced transparency 1 } accounts for the inhomogeneous broadening of the medium normalized to the unit area. \\

The initial conditions are 
\begin{equation}\label{physical boundary conditions}
	\lim_{\tau \rightarrow -\infty} N(\xi=0,\tau,k)=-1,\qquad \lim_{\tau \rightarrow -\infty} f(\xi=0,\tau,k)=0,
\end{equation}
and $\mathcal{E}(x=0,\tau)$ is assumed to decay sufficiently rapidly as $\tau\rightarrow \pm\infty$.\\

The above notations are due to Lamb \cite{PhysRevA.9.422} who was the first one to prove the integrability of the Self-induced transparency.\\

In order to establish a relationship between the set of equations \eqref{Self induced transparency} and the associated initial conditions \eqref{physical boundary conditions} to  Eqs \eqref{full NLS}, \eqref{x evolution of X}, we start with Eq. \eqref{full NLS} with $\alpha=0$:
\begin{equation}\label{our equation}
	q_t=\frac{2}{\pi}\int_{\mathbb{R}}\frac{g(t,k)}{a_0(k)}\Phi_1^+(x,t,k)\Psi^+_1(x,t,k)\mathrm{e}^{\mathrm{i}\lambda \left(HG(t,k)|b_0|^2(k)\right)}dk,\quad x\in\mathbb{R},\quad t>0.
\end{equation}
We recall that $\Phi_1^+$ and $\Psi^+_1$ are entries of the matrices $\Phi$ and $\Psi$, which are the fundamental solutions of the $x$ part of the Lax pair \eqref{x evolution of X}. Following   \cite{doi:10.1063/1.1666551}, we make the following identifications:
\begin{subequations}\label{how to map}
	\begin{align}
		&x=\tau, \quad t=\xi, \quad  q=\frac{1}{2}\mathcal{E}, \quad f=\Phi_1^+(x,t,k)\Psi^+_1(x,t,k), \quad k\in\mathbb{R},\\
		&N(x,t,k)=\Phi^+_1(x,t,k)\overline{\Psi^+_2}(x,t,k)+\overline{\Phi^+_1}(x,t,k)\Psi^+_2(x,t,k), \qquad k\in\mathbb{R}.
	\end{align}
\end{subequations}
Finally, in order to complete the mapping, we choose $g(l,t)$ to be time independent i.e. $g(l,t)=g(l)$. This mapping ensures also the consistency of the  initial conditions \eqref{physical boundary conditions} and \eqref{mathematical boundary conditions} provided
\begin{equation}\label{transparency}
	b_0=0.
\end{equation}
Hence, the relation $|a|^2-\lambda|b|^2=1$ implies $|a_0|=1$, which is consistent with the terminology \lq\lq{}self-induced transparency\rq\rq{}. \\

The simplest solution associated to the system \eqref{Self induced transparency} under the condition  \eqref{transparency}, is the one-soliton solution corresponding to a single bound state $k=k_1$ of the eigenvalue problem \eqref{x evolution of X} in the upper half $k$-plane. This corresponds to a  zero of  $a_0$. The zeros of  $a_0$ as well as soliton solutions are discussed in the appendix B. In the original physical frame $(\xi,\tau)$, the one-soliton solution of \eqref{Self induced transparency} is given by:
\begin{equation}\label{self transparency soliton}
	\mathcal{E}=
	-4\mathrm{i}\frac
	{\overline{c_1}
		\mathrm{e}^
		{-2\mathrm{i}\overline{k}_1\tau-2\mathrm{i}\overline{k}_1^2\xi+\mathrm{i}HG_1}
	}
	{1+
		\frac
		{|c_1|^2}
		{4{(k
				_{1\text{I}})
				^2}}
		\mathrm{e}^
		{-4k_{1\text{I}}\tau-8k_{1\text{R}}k_{1\text{I}}\xi
		}
	}.
\end{equation}
The velocity of the soliton  \eqref{self transparency soliton} is proportional to  $-1/k_{1\text{I}}$, where $k_1=k_{1\text{R}}+\mathrm{i}k_{1\text{I}}$, $k_{1\text{I}}>0$.  In the co-moving frame $(\xi,\tau)$ with the incident  laser pulse, the negative speed of the soliton means that it is propagating in the same direction as the incident pulse, but  moves with a speed slower than the  speed of light. The amplitude of the soliton is  proportional to  the magnitude of $k_{1\text{I}}$, whilst its width depends on  the choice of the broadening profile $g(k)$ in \eqref{our equation}.\\

Self-Induced Transparency and its possible extensions \cite{Ginzburg2021SelfInducedTS} are important examples of physically significant cases where the nonlinearity and the dispersion are both absent in the time evolution equation. This example shows that  systems of this type, belong to the special case  of $ \alpha =0 $ of the present formalism, and as shown in Eq \eqref{truncated equation}, are integrable.\\

\noindent \underline{Stimulated Raman Backscattering}\\

To initiate the process of the so-called stimulated Raman Backscattering, a small Gaussian initial seed pulse and a constant initial pump with a sharp front, are injected in a plasma in opposite directions.
The mismatch between the seed and the pump frequencies causes the formation of a Langmuir wave propagating in the opposite direction of the pump (backscattering) depleting the pump. The fraction of the pump energy contained in the leading spike is determined by the initial amplitude of the seed pulse integrated over its width \cite{PhysRevLett.82.4448}. Once the pump depletion begins, the leading amplified spike propagating directly behind the seed pulse grows and contracts. \\

This phenomenon, which constitutes a one-dimensional resonant quasi-static 3-wave interaction, together with the lowest order relativistic nonlinearity and group velocity dispersion terms, can be described by the following dimensionless set of equations \cite{PhysRevE.93.063210}:
\begin{subequations}\label{SBS}
	\begin{align}
		&\mathcal{A}_{\xi}=-\mathcal{B}\mathcal{F}, \\
		&\mathcal{F}_{\xi}=\mathcal{A}\overline{\mathcal{B}}-\mathrm{i}\Gamma \mathcal{F},\\
		&\mathcal{B}_{\tau}=\mathcal{A}\overline{\mathcal{F}}-\mathrm{i}s\mathcal{B}_{\xi\xi}+\mathrm{i}|\mathcal{B}|^2\mathcal{B},
	\end{align}
\end{subequations}
where $\mathcal{A}$, $\mathcal{B}$ and $\mathcal{F}$ are the envelopes of the pump pulse, counter-propagating shorter seed pulse, and resonant Langmuir wave, respectively. The independent variable  $\tau$ measures the elapsed amplification time, whilst $\xi$ measures the distance from the original seed front. The parameter $s$, which characterises the group velocity dispersion of the amplified pulse, depends only on the ratio of the plasma to laser frequency. In  a strongly under-critical plasma, $s\ll 1$ and in near critical plasmas, $s\gg 1$. The parameter $\Gamma$ which specifies the rescaled detuning, is proportional to $\delta\omega=\omega_\mathcal{F}+\omega_\mathcal{B}-\omega_\mathcal{A}$, where $\omega_\mathcal{F}$, $\omega_\mathcal{B}$, $\omega_\mathcal{A}$ are the frequencies of the pump, seed, and Langmuir waves, respectively.\\ 

In order to map Eqs \eqref{SBS} to our equations \eqref{full NLS} and \eqref{x evolution of X}, we start by setting
\begin{align}\label{SBS mapping 1}
	x=-\xi, \quad t=-\tau, \quad s=\frac{1}{2}, \quad \lambda=-1, \quad \alpha=1, \quad \Gamma=2k,\quad k\in \mathbb{R}.
\end{align}
Then, taking into account the fact that the pump pulse is not reflected (but depleted), we have the simplifying situation of $b_0=0$ and consequently, using $|a|^2-\lambda|b|^2=1$,  we can choose $a_0=1$. Hence, replacing these specific values of $ a_0 $ and $ b_0 $ in Eq \eqref{phi psi related}, relating the two fundamental eigenfunctions of \eqref{x evolution of X} through the $S$ matrix \eqref{new S matrix}, taking into account the symmetry relations \eqref{matrix elements},  and restricting $k$ to belong to $\mathbb{R}$, the mapping is completed by setting
\begin{align}\label{SBS mapping 2}
	\mathcal{B}=-q, \qquad \mathcal{A}=\phi^+_1, \qquad \overline{\mathcal{F}}=\psi^+_1,  \qquad g(l)=\frac{\pi}{2}\delta(l-k),\qquad l\in\mathbb{R}, 
\end{align}
along with the natural initial conditions:
\begin{align}\label{SBS initial conditions}
	b_0(k)=0, \qquad  a_0(k)=\lim_{x\rightarrow -\infty}\Psi^+_{02}(x,k)=1, \qquad k\in\mathbb{R}.
\end{align}

In the case that  the eigenvalue problem \eqref{x evolution of X} possesses a single bound state at $k=k_1$ in the upper half $k$-plane, the envelop of the seed pulse is a soliton, which in the frame $(\xi, \tau)$, is given by (appendix B)
\begin{equation}\label{SBS soliton}
	\mathcal{B}=
	2\mathrm{i}\frac
	{\overline{c_1}
		\mathrm{e}^
		{2\mathrm{i}\overline{k}_1\xi+2\mathrm{i}\overline{k}_1^2\tau+\mathrm{i}HG_1}
	}
	{1+
		\frac
		{|c_1|^2}
		{4{(k
				_{1\text{I}})
				^2}}
		\mathrm{e}^
		{4k_{1\text{I}}\xi+8k_{1\text{R}}k_{1\text{I}}\tau
		}
	}.
\end{equation}

The resonant interaction of wave packets in nonlinear media was first shown to be integrable, in  its most general form, in \cite{SRBS_Zakharov_Manakov}. It is worth noticing that one of the important results in the mentioned work is expressed as follows: \lq\lq{}the physical picture of the interaction depends in a fundamental way on the ratio of the velocities of the pumping
and the secondary waves\rq\rq{}. The novelty of the present work, is that the mapping \eqref{SBS mapping 2} allows us, to determine  the exact plasma parameter, related to this \q{ratio}, for which the Stimulated Raman Scattering is integrable. Namely, for $s=\frac{1}{2}$, i.e. at the \q{middle} of the under-critical or near critical stages.

\section{Conclusions}
The main achievements of this work is summarized by eqs \eqref{full NLS}-\eqref{t evolution of X}.\\   

The methodology elaborated on here can be applied to any integrable PDE in one spatial dimension (equations in 1+1). We hope that this work will motivate researches in the area of integrable systems to construct physically interesting forced versions of other integrable equations. In this connection it will be intresting to investigate symmetries, conversation laws, and integrable hierarcies associated with forced integrable PDEs.\\

It is important to emphasize that  this new methodology can also be extended to integrable evolution PDEs in two spatial dimensions (equations in 2+1). In this connection, it is noted that for the implementation of the dressing method to 2+1, the RH problem is replaced by either a nonlocal RH problem \cite{Fokas_Zakharov_1992} or a nonlocal d-bar problem \cite{Zakharov1985}. For equations in 2+1, the operator $\Delta_g$ is replaced by the operator $\Delta_2$ defined in \cite{FOKAS2022128290}.\\

In the early works on nonlinear evolution equations involving singular dispersion relations \cite{Leon1990b}, \cite{Latifi-Leon}, it was thought that the IST method cannot be systematically applied to nonlinear evolution equations with forcing. Therefore, the authors of these early works, related these equations to a RH problem, attempted to solve them through the d-bar approach, and {\it did not} seek for the associated Lax pair. Further progres was made in \cite{doi:10.1063/1.529443}. V.K. Melnikov disagreed with the point of view expressed in  \cite{Leon1990b}-\cite{doi:10.1063/1.529443} and published a series of papers attempting to prove that the IST method can be applied to forced nonlinear evolution equations \cite{Melnikov_1990}, \cite{Melnikov_1992}. However in his works it is almost impossible to identify the corresponding Lax pairs. Since then, a large  number of publications, studying nonlinear evolution equations with a self-consistent source, can be found in the literature. Among these works we mention the followings:
the KdV,  mKdV, NLS, Degasperis-Procesi, KP, dispersionless KP, Dispersionless mKP, Davey Stewartson, Ishimori, Toda Lattice and Heisenberg ferromagnetic equations with a self-consistent source, are studied in  \cite{Zeng_2003}, \cite{Yao_2008}, \cite{Shao_2005},  \cite{Huang_2008}, \cite{doi:10.2991/jnmp.2006.13.2.4},  \cite{Xiao_2006}, \cite{SHEN2009585}, \cite{doi:10.1063/1.3127482},  \cite{Liu_2005}, \cite{doi:10.1142/S0219887815501340}, respectively. It should be noted that the associated Lax pairs are not exhibited in some of these papers. In each of these cases, the self consistent source take various forms, such as $\sum_i\psi_i\phi_{ix}$, $\sum_i(\psi_i\phi_{i})_x$, $\sum_i(\psi_i)^2$, $\sum_i\psi_i\phi_{i}$ where, $\phi_i$ and $\psi_i$ are eigenfunctions of the associated spectral operator. However, it is not clear which particular solutions of the associated spectral problems should be used. Moreover, there is no systematic procedure of finding the Lax pair. Furthermore, the approach used in the present work proves, as an immediate consequence,  the integrability of the
physically significant cases similar to those captured by Eq \eqref{truncated equation},  where  both $ Q^3\sigma_3 $, i.e. the nonlinearlity, and $Q _{xx}\sigma_3 $, i.e. the dispersion, vanish in the evolution equation \eqref{solution of RH}. Indeed, from a physical point of view, as illustrated in the case of self induced transparency \eqref{Self induced transparency}, the set of equations \eqref{x and t evolution for mu}-\eqref{truncated equation} represents the general case of the three waves interaction where, the frequency's missmach of two of the three waves, induces the growth and propagation of the third wave.

\setcounter{equation}{0}
\numberwithin{equation}{section}
\appendix

\section*{Appendix A}\label{A}

\setcounter{section}{1}
\newtheorem{Proposition}{Proposition}[section]

In what follows we first examine what happens if a given Lax pair is modified by the addition of the operator $N$ to its $t$-part.\\
\begin{Proposition}
	The compatibility condition of the Lax pair
	\begin{subequations}\label{modified Lax pair}
		\begin{equation}
			\psi_x+N_1\psi=0,\label{modified x part}
		\end{equation}
		\begin{equation}
			\psi_t+N_2\psi+N\psi=0,\label{modified t part}\
		\end{equation}
	\end{subequations}
	is the equation
	\begin{equation}\label{modified evolution equation}
		(N_{1t}-N_{2x}-[N_1,N_2])\psi=(N_x+[N_1,N])\psi.
	\end{equation}
	Furthermore, the following identity is valid:
	\begin{equation}\label{additional relation}
		(\psi\sigma_3\psi^{-1})_x=-[N_1,\psi\sigma_3\psi^{-1}].
	\end{equation}
\end{Proposition}
\begin{Dem}
	Equation  \eqref{modified evolution equation} follows from the usual commutativity condition by replacing $ N_2 $ by $ N_2+N $.\\
	
	Equation \eqref{additional relation} can be verified via elementary computations. For example,
	\begin{equation}\label{proof the relation}
		(\psi\sigma_3\psi^{-1})_x=\psi_x\sigma_3\psi^{-1}-\psi\sigma_3\psi^{-1}\psi_x\psi^{-1}.\nonumber
	\end{equation}
	Replacing $\psi_x$ by $-N_1\psi$ we obtain \eqref{additional relation}. \qquad QED\\
	
	For the forced NLS,
	\begin{equation}\label{ choice for NLS}
		N_1=\mathrm{i}k\sigma_3-Q,\qquad N=\frac{1}{2\mathrm{i}}\left(Hg\psi_x\sigma_3\psi^{-1}\right).
	\end{equation}
	Thus,
	\begin{align}
		N_x=&\frac{1}{2\mathrm{i}}\left(Hg(\psi\sigma_3\psi^{-1})_x\right)=-\frac{1}{2\mathrm{i}}\left(Hg[\mathrm{i}k\sigma_3-Q,\psi\sigma_3\psi^{-1}]\right)
		\nonumber\\
		&=-\frac{1}{2}[\sigma_3,\left(Hkg\psi\sigma_3\psi^{-1}\right)]+[Q,N].
		\nonumber
	\end{align}
	Using the identity
	\begin{equation}\label{the latest identity}
		\left(Hkf\right)(k)=k\left(Hf(k)\right)+\frac{1}{\pi}\int_{\mathbb{R}}f(l)dl,
	\end{equation}
	we find
	\begin{equation}\label{full evolution equation}
		N_x+[N_1,N]=-\frac{1}{2\pi}\left[\sigma_3,\int_\mathbb{R}g\psi\sigma_3\psi^{-1}dl\right]=\frac{1}{\pi}\int_\mathbb{R}g(\psi\sigma_3\psi^{-1})_Odl\sigma_3.
	\end{equation}
	Thus, the Lax pair \eqref{modified Lax pair} with $ N_1 $, $ N $  given in \eqref{ choice for NLS}   and
	\begin{equation*}
		N_2=\alpha \left[ kQ- \frac{\mathrm{i}}{2}\left(  Q_x+Q^{3}  \right) \sigma_3   \right],
	\end{equation*}
	yields
	\begin{equation*}
		Q_t+\frac{\mathrm{i}\alpha}{2}Q_{xx}\sigma_3-\mathrm{i}\alpha Q^{3}\sigma_3=\frac{1}{\pi}\int_{ \mathbb{R}}g\left(\psi\sigma_3 \psi^{-1} \right)_Odl\sigma_3,
	\end{equation*}
	which is eq. \eqref{solution of RH}. 
\end{Dem}

\section*{Appendix B}\label{B}
\stepcounter{section}
In the case of $\lambda=-1$, the RH problem for $\mu$ may be singular, namely, the function $a_0(k)$ may have zeros. There exists a well known procedure which maps the singular RH problem to a regular one supplemented with a system of linear algebraic equations \cite{doi:10.1137/0527040}. Alternatively, one may employ the machinery of Darboux transformation \cite{Chvartatskyi2016}\cite{Shao_2005}. In what follows, we implement this procedure to the case that $a_0(k)$ has a single zero at $k=k_1$,
\begin{equation}\label{k}
	k_1=k_{1\text{R}}+\mathrm{i}k_{1\text{I}}, \qquad k_{1\text{I}}>0.
\end{equation}
Let $M$ be defined in terms of $\mu$ via the equation
\begin{equation}\label{definition of M}
	\mu = (kI+B)M
	\begin{pmatrix}  \frac{1}{k-k_1} & 0 \cr 0 & \frac{1}{k-\overline{k_1}} \end{pmatrix},
\end{equation}
\noindent where  $B$ is a $k$-independent matrix and $M$ is free of poles. Then, the first column of $\mu$ has indeed a pole at $k=k_1$ and the second column has a pole at $k=\overline{k_1}$. Also $M$, like the function $\mu$, tends to $I$ as $k\to\infty$. The transformation \eqref{definition of M} and Eqs. \eqref{jump J} imply that the function $M$ satisfies a regular RH problem with a jump given by
\begin{equation}
	M^-=M^+\begin{pmatrix} 1 &  -\frac{b_0(k)}{\overline{a_0}(k)}\frac{k-k_1}{k-\overline{k_1}}\mathrm{e}^{-\mathrm{i}\theta}\cr - \frac{\overline{b_0}(k)}{a_0(k)}\frac{k-\overline{k_1}} {k-k_1}\mathrm{e}^{\mathrm{i}\theta}& \frac{1}{|a_0(k)|^2} \end{pmatrix}, \quad \theta=2kx+2k^2t-HG.
\end{equation}
The residue conditions imply the following equations for the entries of the matrix $B$:
\begin{subequations}\label{algebraic}
	\begin{align}
		(k_1I+B)M\begin{pmatrix}1 \cr -d_1 \end{pmatrix}=0,\\
		(\overline{k_1}I+B)M\begin{pmatrix}\overline{d_1} \cr 1 \end{pmatrix}=0,
	\end{align}
\end{subequations}
where 
\begin{equation}
	d_1=\frac{c_1}{k_1-\overline{k_1}}\mathrm{e}^{-\mathrm{i}\theta_1}, \qquad \theta_1=2k_1x+2k_1^2t-HG_1, \qquad G_1=G(k_1,t).
\end{equation}
In the case of a pure soliton, $b_0=0$ and $M=I$. Thus, Eqs \eqref{algebraic} imply,
\begin{equation}
	B_{12}=\frac{k_1-\overline{k_1}}{1+d_1\overline{d_1}}\overline{d_1}=
	-\frac
	{\overline{c_1}
		\mathrm{e}^
		{-2\mathrm{i}\overline{k}_1x-2\mathrm{i}\overline{k}_1^2t+\mathrm{i}HG_1}
	}
	{1+
		\frac
		{|c_1|^2}
		{4{(k
				_{1\text{I}})
				^2}}
		\mathrm{e}^
		{-4k_{1\text{I}}x-8k_{1\text{R}}k_{1\text{I}}t
		}
	}.
\end{equation}
Equation \eqref{Q} yields
\begin{equation}
	q=2\mathrm{i}B_{12}.
\end{equation}

 



\label{lastpage}
\end{document}